# Anisotropic magnetic and magnetotransport properties in morphologically distinct $Nd_{0.6}Sr_{0.4}MnO_3$ thin films

**R S Mrinaleni [1, 2], E P Amaladass[1, 2*], A. T. Sathyanarayana [1, 2], S Amirthapandian[1,2], Jegadeesan P[1], Pooja Gupta [3, 4], T Geetha Kumary[1, 2], and S. K. Rai [3, 4], Awadhesh Mani[1, 2]**

**[1]Material Science Group, Indira Gandhi Centre for Atomic Research, Kalpakkam, 603102, India**
**[2]Homi Bhabha National Institute, Indira Gandhi Centre for Atomic Research, Kalpakkam 603102, India**
**[3]Synchrotrons Utilisation Section, Raja Ramanna Centre for Advanced Technology, PO RRCAT, Indore, Madhya Pradesh 452013, India**
**[4]Homi Bhabha National Institute, Training School Complex, Anushaktinagar, Mumbai, Maharashtra 400094, India**

***Corresponding author:** edward@igcar.gov.in*

**ABSTRACT**

We investigate the magnetic and magneto-transport properties of nanostructured $Nd_{0.6}Sr_{0.4}MnO_3$ (NSMO) thin films grown on (100) oriented $SrTiO_3$ (STO) substrates. The thin films fabricated using the pulsed laser deposition technique have been found to possess two distinct surface morphologies – granular and nano-rod type. Magnetization measurements have revealed that the films with rod-type morphology exhibit improved in-plane magnetic anisotropy. Magnetotransport studies have revealed that the granular thin films display a characteristic butterfly-shaped low-field magneto-resistive (LFMR) behavior. Furthermore, we investigate the anisotropic magneto-resistive (AMR) phenomenon in the samples and we find that morphology greatly affects AMR. Thin films with rod-type morphology show an enhanced AMR %. Such morphology-dependent tunability in magnetoresistance properties over a wide temperature range is potentially interesting for developing oxide-based sensors and devices.

## 1. INTRODUCTION

Magnetoresistance is a system property characterized by a change in resistivity with the application of a magnetic field. The magnetoresistance ratio ($\Delta R/R(H=0)$) is a highly useful quantity in characterizing materials for application in the field of spintronics, non-volatile memory storage, and information processing[1]. In this context, manganites have been one of the most studied compounds in recent decades due to their enormous magnetoresistance ratio. This phenomenon is termed as the colossal magnetoresistance (CMR) effect which occurs around the magnetic transition in manganites[2–5]. Manganites have the general formula $RE_{1-x}AE_xMnO_3$ (where, RE= $La^{3+}$, $Nd^{3+}$, $Pr^{3+}$, $Sm^{3+}$, and AE = $Ca^{2+}$, $Sr^{2+}$, Ba,$^{2+}$, etc.) and exhibit a variety of magnetic phases depending on the value of x, which ranges from 0 to 0.9[2–5]. They exhibit charge ordering, orbital ordering, spin-polarisation, phase separation, metal-to-insulator transition (MIT), kinetic arrest, etc[2–5]. These properties are interesting for basic research as well as application purposes.

$Nd_{1-x}Sr_xMnO_3$ is a manganite system with a mid-bandwidth and exhibits properties and magnetic phases similar to those observed in the popular manganite $La_{1-x}Sr_xMnO_3$[2–5]. For x = 0.4, $Nd_{0.6}Sr_{0.4}MnO_3$ (NSMO) shows paramagnetic insulator to ferromagnetic metal transition at 270 K in its bulk form[6]. While many studies have been conducted on its bulk, the properties in the thin film form have been relatively less explored and are only recently gaining interest[7–9]. Thin films of manganites show significant enhancement in MR values due to strain offered by the substrate as compared to their polycrystalline or single-crystal forms[10]. The strain can be tuned by choosing appropriate substrates with low lattice mismatch, by growing thin films on vicinal substrates, and by preferring different substrate orientations[11,12]. Due to the strong correlation among the spin, charge, and lattice degrees of freedom in manganites, any external perturbation can alter their physical properties to a great extent. For instance, in manganite thin films, the strain induced by the substrate can cause the magnetization easy axis to lie in-plane or out-of-plane depending on whether the strain is tensile or compressive[11,12]. The strain relaxation can cause alterations the to growth mode resulting in different morphology and nanostructuring in the thin film. Thin films grown on vicinal substrates possessing misoriented steps possess unique structural variants due to the strain offered by the stepped surface. The step edges at the substrate surface cause symmetry breaking and unequal strain in the thin films resulting in uniaxial magnetic anisotropy[11,12]. Reports claim that in the case of $SrTiO_3$ (STO) (100) with low miscut ( $< 1^o$), the manganite thin films experience an in-plane tensile strain as a result of which magnetization easy axis lies in-plane along <100> [11,12]. Additionally, the fabrication of manganites in the thin film form that possess enhanced MR is essential for spintronics and information processing applications due to their increased magnetic sensitivity compared with widely used Hall sensors[13]. By optimally tuning the deposition parameters of thin films, one can tailor a variety of physical properties such as a high Curie temperature close to room temperature, a large temperature-coefficient of resistance (TCR), an enhanced magnetic anisotropy, a large low-field magnetoresistance (LFMR) and high anisotropic magnetoresistance (AMR). Previous works have tried to improve the MR property by growing textured LSMO thin films on $Si/SiO_2$, quartz, $Al_2O_3$, and even flexible mica substrates[14–16]. They found that the resulting morphology of the thin films with triangular and polygonal crystallites covering the thin film surface and columnar growths greatly improved the LFMR% and AMR %[13]. Thus, in magnetic thin films, the morphology and the type of nanostructuring present in the system greatly affect its magnetic anisotropy and magnetotransport properties.

In our previous work, we synthesized NSMO thin films with two distinct morphologies – granular and nano-rod type. Both the films were grown on STO substrates possessing miscut and the reason for such nanostructuring was examined in detail[17]. We observed that films with rod-type morphology showed improved crystallinity and enhanced TCR %. In the present work, we investigate these thin films' magnetic and magnetotransport properties. Such investigations are exceptionally useful to tailor the magnetic and magnetotransport properties of manganite thin films for oxide-based devices.

## 2. Experimental methods

The NSMO thin films were fabricated using the PLD technique with a KrF excimer laser source ($\lambda = 248$ nm). The films were deposited on $SrTiO_3$ (STO) substrate with a miscut value ranging between $0.3^o$ to $0.4^o$ at a laser energy density of 1.75 $J/cm^2$ at 3Hz. The thin films were deposited with an oxygen partial of 0.36 mbar and the substrate temperature was fixed at 750 $^oC$. Post deposition, the films were in situ annealed at 750 $^oC$ for 2h followed by another ex-situ annealing at 950 $^oC$ in an oxygen atmosphere for 2h. The detailed description of thin film preparation, morphology, and structural characterization are described elsewhere[17]. The transport properties of the thin films (NS-G and NS-R) have been studied using a cryo-based magneto-resistance setup from Cryogenic UK. Magnetization measurements have been performed in a SQUID-based vibrating sample magnetometer (VSM) system using a Quantum Design EverCool SQUID magnetometer.

## 3. RESULTS AND DISCUSSION:

### *3.1. Morphology, structural, and transport properties of the nanostructured thin films:*

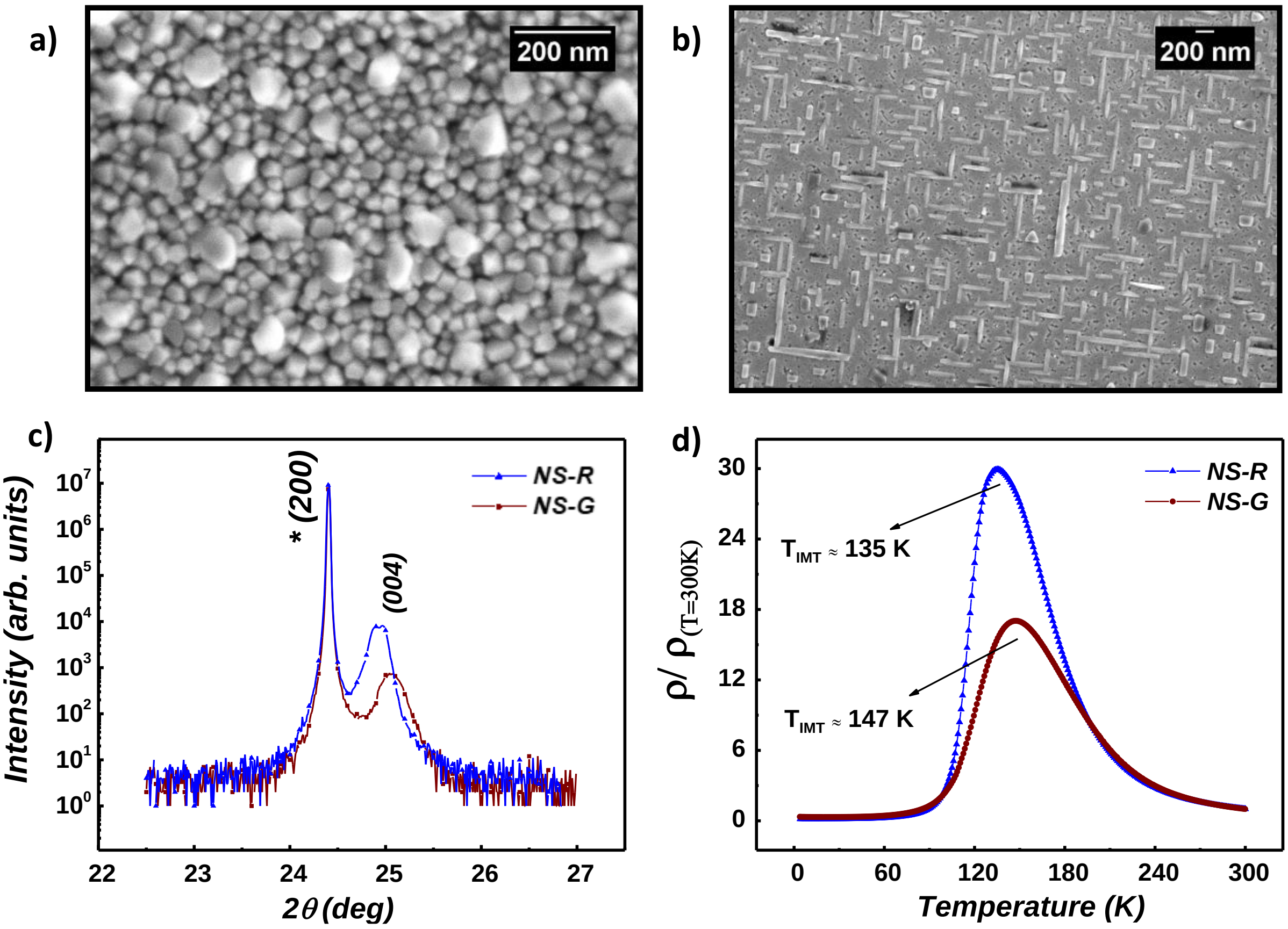

***Figure 1:*** *Scanning electron microscopy (SEM) images of NSMO thin films on STO with a) NS-G - granular morphology. b) NS-R – crossed nano-rod morphology. c) High-resolution XRD scans of NS-G and NS-R thin film (* - STO peaks). d) Normalized temperature-dependent resistivity plots of NS-G and NS-R samples exhibiting metal-to-insulator transition (MIT).*

Figure 1 (a) shows the morphology of the NSMO thin films with a granular morphology and figure 1 (b) shows a nano-rod type morphology in which the crossed nano-rods are embedded in a smooth matrix of NSMO. We will refer to the film with granular morphology as NS-G and one with rod-morphology film as NS-R henceforth. The reason for such nanostructured growth is discussed in detail elsewhere[17]. From XRD investigations, we found that thin films with granular morphology possessed multiple reflections owing to their polycrystalline nature whereas the films with rod-type morphology possessed only out-of-plane reflection[17]. A high-resolution XRD scan around the (200) plane of STO is shown in figure 1 (c) and it gives NSMO (004) reflections along the out-of-plane direction for both films. But we can see that NS-R has reduced full-width half maxima as compared to NS-G with lesser relative intensity. Figure 1(d) shows the normalized temperature-dependent resistivity plots of the NSMO thin films. The MIT of NS-G occurs at 147 K and in NS-R at 135 K. The slope of the metallic transition below

MIT was found to be sharper in the case of NS-R as compared to NS-G. This signature of sharp transition into the metallic phase was found across all our thin films with rod morphology. A detailed description explaining this effect can be found elsewhere[17]. Further, the magnetic and magneto-transport properties of the nanostructured NSMO thin films are studied comprehensively.

### *3.2.Magnetization measurements:*

The zero-field cooled (ZFC) and field-cooled (FC) magnetization of NS-G and Ns-R are shown in Figures 2 (a), and (b) respectively where the measurements were performed at 0.1 T with the field parallel to the sample surface. In Figure 2 (c), and (d) the in-plane magnetization hysteresis is plotted as a function of temperature for NS-G and NS-R, respectively. On comparing the magnetic hysteresis curves of NS-G and NS-R, we find that the hysteresis loop area increases systematically as the temperature approaches 4 K confirming that the samples enter the FM state. The magnetic transition temperature was determined by taking the derivative of the temperature-dependent magnetization curves. The Curie temperature ($T_{Curie}$) of NS-G and NS-R was found to be ~ 218 K and ~ 100 K, respectively. We can also see that the MIT occurs at 135 K in the case of NS-R which is higher than the Curie temperature, which seems counterintuitive. Thus, the $T_{Curie}$ obtained from this method (derivative curve of magnetization) may not be the exact temperature at which the magnetic phase transition occurs. Additionally, the magnetic hysteresis curves of NS-R also show that the sample is not completely in PM state above 100 K up to 150 K. In manganites, the reduction in $T_{Curie}$ is assigned to the presence of oxygen-vacancies/defects in the system, which affects the double-exchange interaction responsible for the magnetic transition[7]. The samples NS-G and NS-R were synthesized under the same PLD conditions and annealed in the same batch. Thus, such large differences in $T_{Curie}$ cannot be assigned solely to oxygen deficiency. We believe that this drastic change in $T_{Curie}$ is due to the change in the morphology and crystallinity of the samples.

In the phase-separated manganite system, crystallographic disorders play an important role in the nucleation of the FM phase during the phase transition. In the granular film, the large number of crystallographic defects, grain-boundaries (GBs), and anti-phase boundaries (APBs) act as pinning as well as nucleation points for the growing FM domains. However, studying the nucleation of magnetic domains requires further experimentation using advanced techniques like magnetic force microscopy (MFM) at low temperatures. In a recent study on magnetic domain wall motion in ferromagnetic $SrRuO_3$ (SRO) thin films, Zahradnik *et al.*[18] verified experimentally that APBs act as domain nucleation centers. This was observed in the SRO thin film possessing multi-variant growth as compared to the single-variant film free of islands and APBs grown on STO substrate. Corroborating this, the granular film NS-G possesses a large density of GBs acting as pinning centers thereby exhibiting a higher coercive field as compared to NS-R. The temperature dependence of the coercive field, $H_c$ for NS-G and NS-R is plotted in Figure 2 (f). As the FM phase fraction decreases with an increase in temperature, a reduction in $H_c$ is observed with increasing temperature.

In addition to the tensile strain offered by the substrate (STO) which favors in-plane magnetization, the NS-R exhibits improved in-plane magnetic anisotropy. From the M-H loop at 4K as shown in Figure 2 (d) a sharp switching in magnetization near $H_c$ is observed in NS-R as evident from the larger slope near $H_c$ as compared to NS-G. The magnetic anisotropy strongly depends on the crystallinity of the system[19,20]. According to the Stoner-Wohlfarth model[20], a single-domain sample exhibits a squared M-H loop when the field is applied along the easy axis. Whereas a polycrystalline sample due to lack of magnetocrystalline anisotropy often exhibits an s-type hysteresis[20]. The M-H hysteresis of NS-G is similar to polycrystalline averaged hysteresis loop[20]. Hence, we deduce that there is increased in-plane magnetic anisotropy in NS-R as compared to NS-G. This substantiates the claim of improved crystallinity in NS-R from HR-XRD studies. Additionally, at low temperatures (4 to 10K) the M-H hysteresis of NS-R shows a kink-like feature which is absent at higher temperatures (see Figure 2 (e)). Such a feature may arise due to the presence of competing magnetic phases with different $H_c$[21].

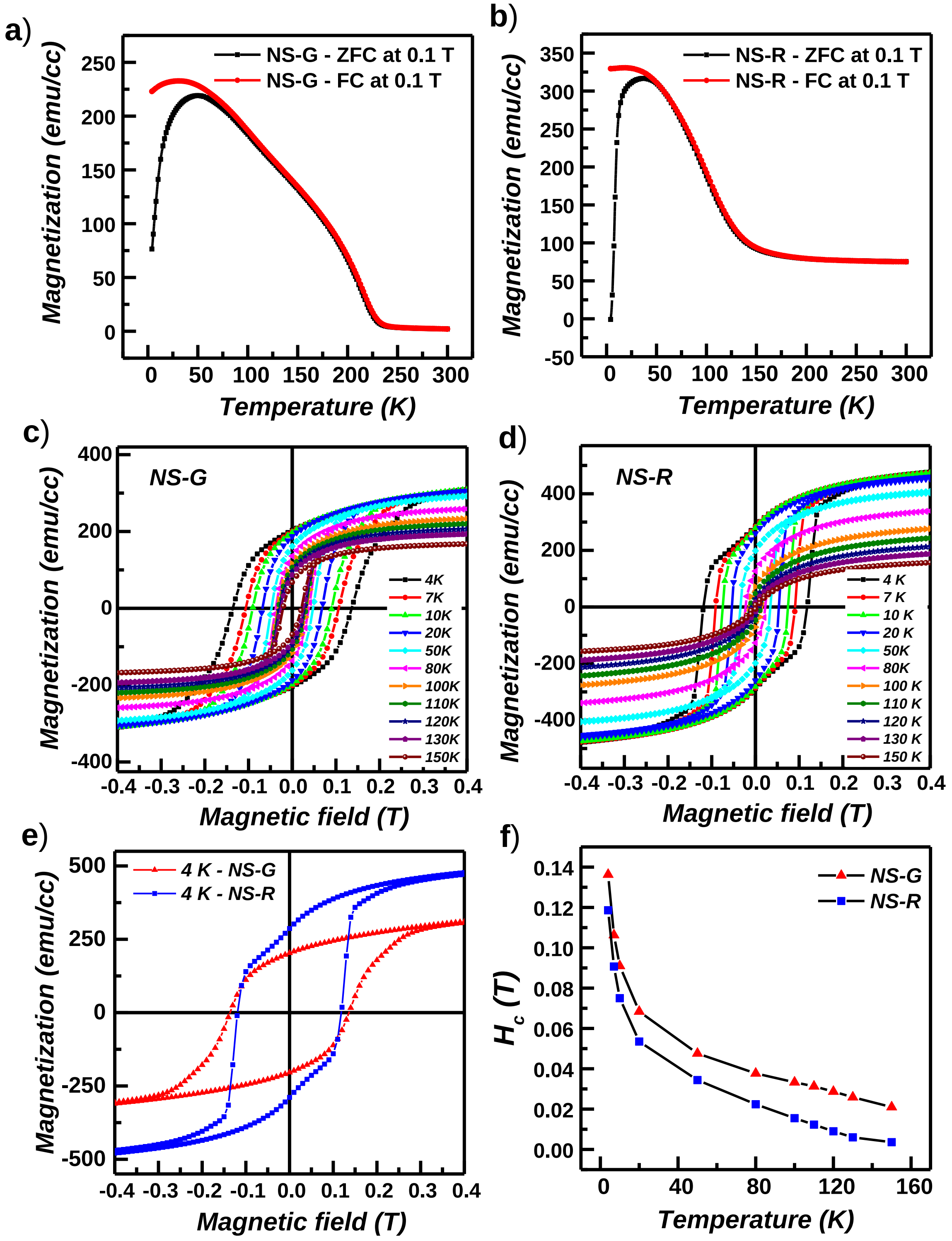

*Figure 2: a), b) Temperature-dependent magnetization curves of NS-G and NS-R, respectively c), d) Magnetization hysteresis curves of NS-G and NS-R at different temperatures e) M-H curve at low temperature (4K) f) Variation of coercive field with temperature.*

### 3.3. Magneto-transport measurements:

In the thin films NS-G and NS-R, the magneto-transport measurements have been carried out in three configurations as shown in Figure 3 and the MR (%) is evaluated using equation (1).

$$\boldsymbol{MR}\ (\%) \ = \frac{\boldsymbol{\rho}(\boldsymbol{H}) - \boldsymbol{\rho}(\boldsymbol{H}=\mathbf{0})}{\boldsymbol{\rho}(\boldsymbol{H}=\mathbf{0})} \times \mathbf{100} \qquad (1)$$

In the G1 configuration, the magnetic field is perpendicular to the sample surface (along Z) and the current direction is along the sample surface (along X). In configuration G2 and G3, the magnetic field (along Z) is parallel to the sample surface. In the case of G2, the direction of current flow is kept perpendicular to the magnetic field (along Y) and in G3, the current is applied parallel to the direction of the field (along Z). The magneto-resistance of the thin films is measured in the hall-bar geometry at different temperatures from 4 K up to 300 K where the field is ramped from 3 T to – 3 T in a full loop through 0 T (shown in Figure 4) in different configurations. Below 100 K, both of the thin films NS-G and NS-R show clear hysteresis in MR due to the FM state. In all three MR configurations, a colossal drop in MR is observed in both the samples near its $T_{MIT}$ (120 K-curve in MR). The magneto-transport behavior in each configuration is discussed sequentially.

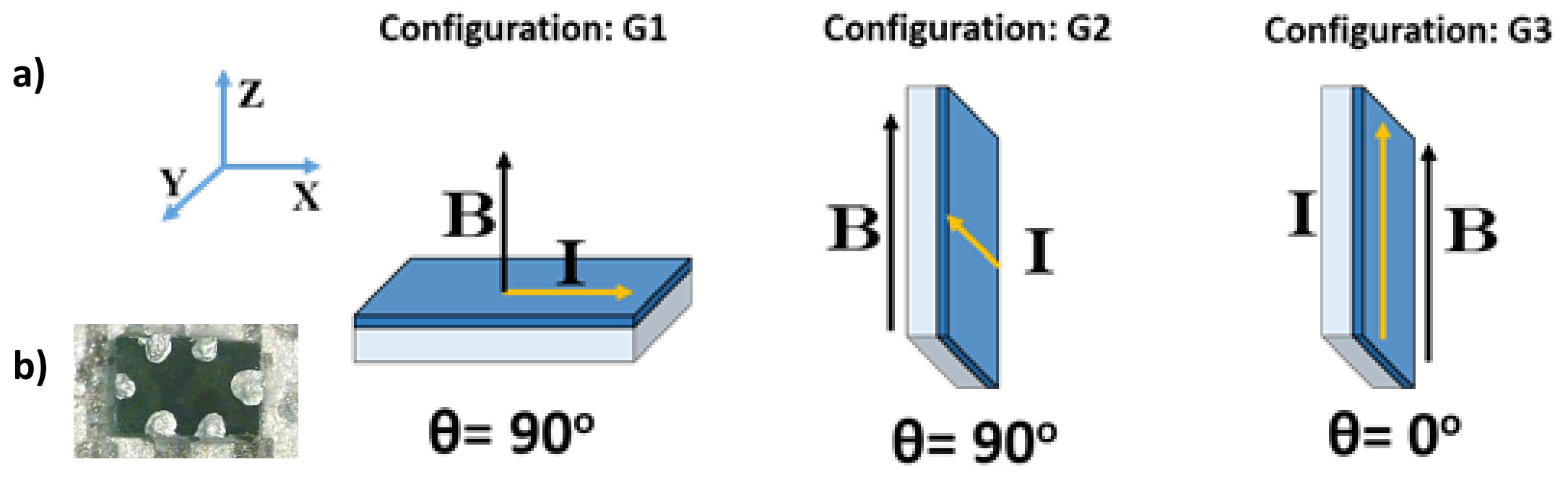

***Figure 3****: a) Schematic of magnetoresistance (MR) measurements in different configurations G1, G2, and G3 on the NSMO thin film (Yellow arrow – Direction of current and Black arrow – Applied magnetic field) b) Electrical contacts on the NSMO thin film for MR measurements.*

In the G1 configuration, a feeble positive MR is observed up to the saturation field as the thin film is not fully magnetized along the field direction. The spin-dependent scattering dominates the electron transport at low fields. Above the saturation field, the spin-dependent scattering reduces with the increase in magnetic field. As a result, both of the thin films display a negative MR in the G1 configuration at B > 1T. At 3T, a maximum CMR % of ~ 87 % is obtained for NS-G whereas ~ 97 % is obtained for NS-R in this configuration.

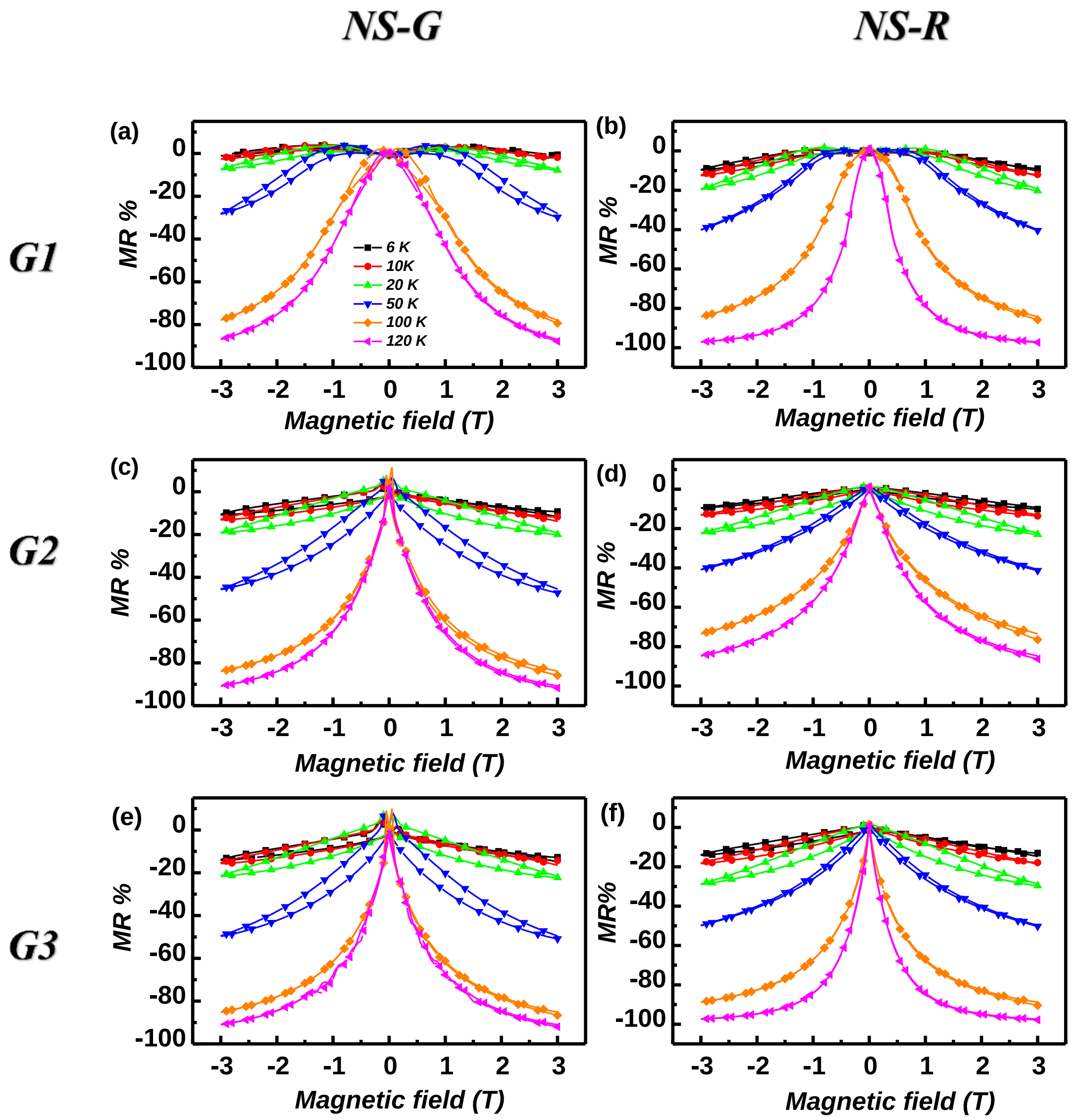

***Figure 4:*** *Temperature-dependent magneto-transport behavior of NSMO thin films. Left column: NS-G; Right column- NS-R; a), b) in G1 configuration c),d) in G2 configuration e),f) in G3 configuration.*

In the G2 and G3 configurations, at 3T, a maximum CMR % of ~ 90 % and 91 % is obtained for NS-G whereas ~ 84 % and 97 % are obtained for NS-R, respectively. In addition to the CMR features at 120K in NS-G, there exists a resistive peak at low fields below $T_{MIT}$. This butterfly-shaped MR / low-field MR (LFMR) is absent in the case of NS-R. Figure 5 gives the MR plot in configuration G3 where the resistive switching (positive MR %) is present in the NS-G thin film throughout the low-temperature regime at low-applied magnetic fields. The butterfly MR/LFMR is very clear at 100K as the field is ramped from 3T to -3T in Figure 5 (c). In the MR plot, the switching field is defined as the position of the peak in the MR curve. The variation of the switching field with temperature shows a similar trend as the coercive field in Figure 5 (f). The $MR_{peak}$ (%) represents MR % at the switching field. Around MIT the resistive switching in NS-G vanishes.

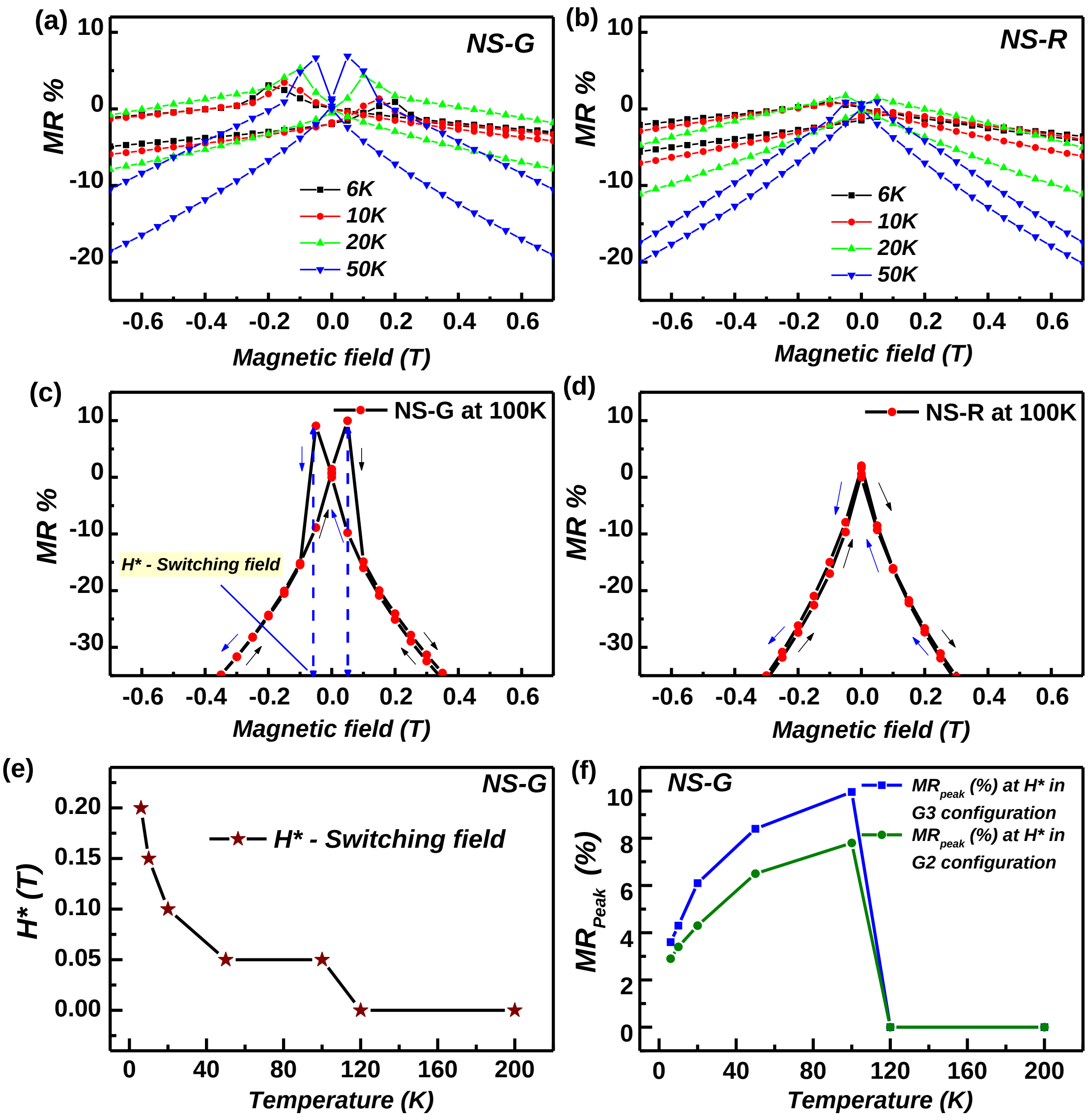

***Figure 5:*** *Low-temperature magneto-transport behavior of NSMO thin films measured in G3-configuration. a), c) A butterfly-shaped MR at low fields in NS-G b),d) Absence of butterfly-shaped MR in NS-R e) Variation of switching field with temperature f) Temperature dependence of the resistive peak ($MR_{peak}$ %).*

The phenomena of LFMR have an origin related to disorder (GBs or APBs) and has been reported in the literature in several manganite systems[22,23,24,25,13] as well as other FM systems[13,26]. The butterfly-shaped MR/LFMR arises in the FM state. Due to the scattering of the spin-polarized transport of electrons across the GB/APBs of FM grains, a resistive peak may be observed. This is because at low fields the spins are not fully aligned and the variation of spins across the boundary (GB/APBs) of two FM grains is large. As the field strength increases, a large number of spins are aligned along the field direction similar to the saturation field, which suppresses the spin scattering effects of the electron transport across the boundary. Thus, a striking LFMR % of up to 10% is observed in the thin film with granular morphology (NS-G). Such effects are absent in thin films with rod morphology (NS-R), due to their morphology with improved crystallinity and reduced GB/APB defects.

Further, as we compare the CMR % of NS-G and NS-R measured in the 3-different MR configurations, the value of CMR% varies. In NS-G, at temperatures below 120 K, the MR % significantly differs across G2 and G3 configurations as shown in Figure 5(f). In NS-R thin films the difference in CMR % between the three different MR configurations is considerably large as seen in Figure 4. This is attributed to AMR phenomena.

The AMR phenomenon is observed in FM systems when the resistance changes depending on the direction of magnetization with respect to the direction of current in the system[27]. It is understood that AMR arises in FM systems due to spin-orbit interaction and additionally the magnetic anisotropy in the system also greatly influences the AMR phenomenon[28]. Also, unlike the 3d-FM systems where AMR % increases with decreasing temperature $T < T_{Curie}$, in the strongly correlated manganite systems an enhanced AMR% is observed near its MIT[29]. This non-monotonic behavior of AMR with respect to temperature, as well as magnetic field, is present across all the manganites. Such behavior is understood to exist due to the electronic non-uniformity near MIT[29]. The AMR ratio is evaluated in-plane and out-of-plane, with contributions from the Lorentz MR and the magnetocrystalline anisotropy but the magnitude of Lorentz MR is trivial[28]. Studies show that AMR features are enhanced in single crystals and epitaxial thin films as compared to bulk due to reduced spin-dependent scattering across GBs[30]. Also, in the case of thin films the substrate imposes a strain resulting in the suppression of Jahn-Teller (JT) distortions owing to an increased AMR %[28]. The diverse aspects of AMR in manganite thin films have been studied across many phase-separated manganite systems including thin films in the literature[29,30,31,32].

$$AMR\,\% = \frac{\rho_{\parallel}^{in} - \rho_{\perp}^{in}}{\rho_o} \times 100 \qquad (2)$$

We have evaluated in-plane AMR % (will be referred as AMR% hereafter) using equation (2) where $\rho_{\parallel}^{in}$ is the resistivity measured, when the field is parallel to current direction (G3 configuration ), $\rho_{\perp}^{in}$ is the resistivity measured, when the field is perpendicular to current direction with the field parallel to the thin film surface (G2 configuration ) and $\rho_o$ is the resistivity measured when H= 0 T. The variation of AMR % with respect to the magnetic field at different temperatures is plotted as contour plots in Figure 6 (a) and (b) along with the line plots in Figure 6 (c) and (d) for thin films NS-G and NS-R respectively. At low temperatures up to 50 K, both the samples NS-G and NS-R show AMR effects in which AMR % increases at low fields and saturates as the field strength is increased. This is because at low temperatures the manganite system is in the FM phase and low fields enhance the FM character of the system.

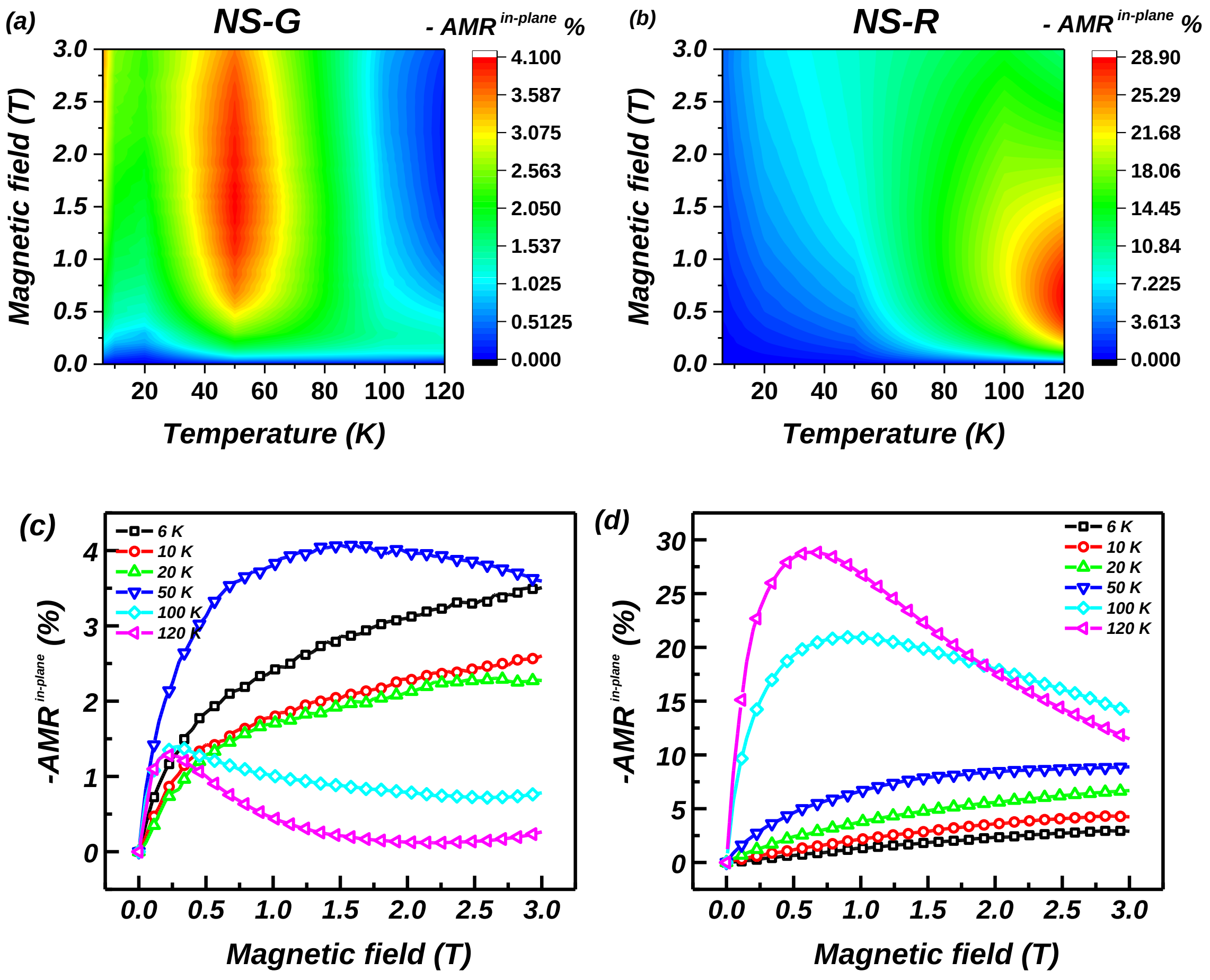

***Figure 6:*** *Variation of AMR % with magnetic fields at temperatures below MIT in a) NS-G and b) NS-R thin film c),d) Line-plot of AMR % evaluated at different temperatures for NS-G and NS-R respectively.*

But at high magnetic fields, the magnetic homogeneity increases. This causes a reduction in the spin-dependent scattering effects at high fields, due to which the AMR curve saturates. In the case of NS-R, as the temperature increases, there is a systematic increase in the AMR% up to MIT, whereas, in NS-G, the increase in AMR % is not systematic. In Figure 6 (a) the contour map shows that at low temperatures T < 20 K the AMR % increase (intensity increases towards red) whereas in Figure 6(b) no such increase in intensity is observed at T < 20K. This low-temperature behavior can arise due to complex mechanisms which may be due to the kinetic arrest of magnetic phases. It would be interesting to study the mechanism of kinetic arrest across these nanostructured systems in the future.

Near MIT temperatures (120 K and 100 K), the AMR % increases up to a certain magnetic field and reduces further. This trend is observed in both samples but is more prominent in the case of NS-R. It is found that the value of the field above which AMR decreases is close to the saturation field as estimated from M-H hysteresis in Figure 2(c) and (d) at 100K and 120K. The large enhancement of AMR is due to increased electronic in-homogeneity near MIT. The sparse fraction of FM phases form percolated paths near $T_{MIT}$ and the application of a magnetic field increases the magnetization of the FM domains giving rise to an initial increase in AMR %. Further, the reduction in AMR % upon increasing the field can be attributed to the co-existing PM phases (FM and PM phases coexist near MIT) which remain unsaturated. This is a dominant effect, in addition to the reduced spin-dependent scattering due to completely saturated FM phases[29,27]. The interplay among the above-mentioned effects is greatly influenced by the morphology of thin films, as the growth/pinning of FM domains is prominently influenced by the morphology (as discussed in section 2.3). As a result, though the granular samples possess AMR, the effects of AMR get averaged due to its polycrystalline nature, and a maximum AMR% of only 4 % is obtained in the NS-G thin film. In addition to other physical properties such as improved crystallinity with reduced defects/GBs and improved in-plane magnetic anisotropy the NS-R thin film with the rod morphology exhibits a maximum AMR of up to 30% at 120 K.

Thus, the non-monotonic behavior of AMR in the manganite thin film is studied thoroughly across the two different nanostructures. In addition to spin-dependent scattering across GBs, magnetic anisotropy and crystallinity have a significant impact on AMR. An enhanced AMR effect arises in the thin film with rod-type morphology. Consequently, such self-organized nanostructures are very useful due to their enhanced physical properties and can be employed for device applications that rely on the AMR phenomena.

## 4. CONCLUSION:

To conclude, the PLD-grown NSMO thin films with granular and nano-rod type morphologies possess diverse magnetic and magnetotransport properties. Magnetization measurements showed that the films with rod-type morphology exhibit improved in-plane magnetic anisotropy as compared to the granular thin film. Magnetotransport measurements revealed that thin films with granular morphology exhibit a striking LFMR of up to ~ 10 % whereas the thin film with rod morphology shows no LFMR. The reason for such LFMR effect is discussed from the point of view of spin-polarized transport of electrons across the grain boundaries in the granular sample. The granular thin film exhibited a maximum CMR% of 91% in the G3 configuration and 90% in the G2 configuration at 3 T. In comparison, for samples with rod morphology, a high CMR% of up to 97% in the G3 configuration and 84% in G2 configuration was obtained. Further, magneto-transport measurements performed in different configurations showed a significant enhancement in AMR % by an order of magnitude in the sample with rod morphology as compared to the granular sample. The reason for this enhancement is analyzed from the point of view of magnetic phase transition. This work emphasizes the importance and influence of nanostructures in tailoring the magnetic and magneto-transport properties of the thin films, especially in the case of manganites which are promising for applications in future oxide-based multi-functional devices.

**Acknowledgments**

One of the authors (R S Mrinaleni) would like to acknowledge the funding support from the Department of Atomic Energy, India. We are grateful to UGC-DAE CSR, Kalpakkam node for the magnetic characterization and measurements of our thin films. The authors are grateful to RRCAT and UGC-DAE-CSR, Indore for the use of beamline facilities.

**Author contributions**

The division of work is as follows: NSMO thin film samples were prepared by R.S.M. SEM imaging was carried out by S.A, J.P. XRD measurements were carried out by PG, S.K.R. Magnetization measurements were carried out by A.T.S. Magneto-transport measurements were carried out by R.S.M and E.P.A. Analysis were done by R.S.M, E.P.A. The writing was carried out by R.S.M. All authors discussed the results and commented on the manuscript. E.P.A., T.G.K., and A.M. supervised this research work.